\documentclass[english]{svjour3}
\usepackage{color}
\usepackage{array}
\usepackage{booktabs}
\usepackage{units}
\usepackage{url}
\usepackage{multirow}
\usepackage{amsmath}
\usepackage{amssymb}
\usepackage{graphicx}
\usepackage{esint}

\makeatletter

\providecommand{\tabularnewline}{\\}

\RequirePackage{fix-cm}

\smartqed  

\makeatother

\usepackage{babel}
\begin{document}

\title{Total spectral distributions from Hawking radiation}

\author{Bogus\l aw Broda}

\institute{Department of Theoretical Physics, Faculty of Physics and Applied
Informatics, University of \L \'{o}d\'{z}, 90-236 \L \'{o}d\'{z},
Pomorska 149/153, Poland.\\
e-mail: bobroda@uni.lodz.pl}

\date{\textcolor{black}{Received: date / Accepted: date}}
\maketitle
\begin{abstract}
Taking into account time-dependence of the Hawking temperature and
finite evaporation time of the black hole, total spectral distributions
of the radiant energy and of the number of particles have been explicitly
calculated and compared to their temporary (initial) blackbody counterparts
(spectral exitances).
\end{abstract}

\textcolor{black}{
}

\section{Introduction\label{sec:Introduction}}

One of the most famous features of the Hawking black hole (BH) radiation
is its blackbody thermal (Planckian) spectrum \cite{Hawking1974,hawking1975particle,Page1976,Page2005,Harlow2016}.
However, on the other hand, it is also well-known, that the BH blackbody
thermality is modified in a number of ways \cite{visser2015thermality,Dvali2016}.
Actually, the shape of the BH spectrum is only approximately Planckian,
with certain key modifications, especially for small BH's. A direct
consequence of the Hawking radiant flux from the BH is its evaporation
\cite{Hawking1974,hawking1975particle}, i.e., the mass $M$ of the
BH monotonically decreases in time. Moreover, since the BH blackbody
temperature $T_{\textrm{H}}$ (the \emph{Hawking temperature}) depends
on the reciprocal of $M$ (see (\ref{eq:TeqTH})), $T_{\textrm{H}}$
is a growing function of time \cite{Hawking1974,hawking1975particle}.
The author of \cite{visser2015thermality} has presented an elegant
and exhaustive discussion of various modifications and limitations
of the thermality of the Hawking radiation. According to \cite{visser2015thermality}
the Planckian shape of the Hawking spectrum will be modified by at
least three distinct physical effects: greybody factors, adiabaticity
constraints and available phase space. From our perspective, somewhat
arbitrarily, we could add to the list the following (not quite independent,
nor new) modifications of the thermality: the BH radiation lasts for
a finite period of time $t_{\textrm{e}}$ (the \emph{evaporation time})
\cite{Hawking1974,hawking1975particle}, and the total number $N$
of the particles emitted by the BH is finite. Interestingly, it appears
that the particles are being emitted from the BH (in a sense) rather
rarely \cite{gray2016hawking}. The mentioned (finite) total number
$N$ (strictly speaking, its average) of particles (photons) emitted
has been calculated quite recently in \cite{alonso2015entropy}, and
especially in \cite{muck2016hawking}. The finiteness of $N$ is not
quite unexpected, but it is not obvious for gapless particles, e.g.\ for
photons. Moreover, this fact (i.e., the finiteness of $N$) could
possibly also influence discussion around the BH information paradox
\cite{Harlow2016}. Since we deal with a finite process ($t_{\textrm{e}}<+\infty$),
and the Hawking temperature $T_{\textrm{H}}$ is a (non-constant)
function of time, what could be physically more representative for
the BH evaporation process than the initial or temporary blackbody
spectrum expressed by the spectral radiant exitance $e_{\omega}$
(see (\ref{eq:eomega})) is the total spectral distribution of energy
$E_{\omega}$ coming from the total particle production from the BH.
Therefore, the aim of the present work is to calculate the total spectral
distribution of energy $E_{\omega}$, and also its actinometric counterpart
$N_{\omega}$ (the\emph{ }total spectral distribution of the number
of particles), both emitted during the whole BH evaporation time $t_{\textrm{e}}$.
In other words, we assume complete evaporation (the final mass $M_{\textrm{f}}=0$).
This assumption implicitly ignores possible modifications of the evaporation
process for the temporary mass of the order of the Planck mass, when
unknown quantum gravity effects should probably be taken into account.
For simplicity, we also assume that at any instant of time the radiation
is given by the Hawking blackbody formula for the Schwarzschild BH.
Moreover, we limit our discussion to one species of massless particles,
i.e., to photons. Thanks to these simplifications, we are able to
express the both formulas in closed analytical forms as polylogarithm
functions. The both calculated distributions are not only interesting
in itself, but also they give rise to the approximate notion of the
(time) average(d) temperature $\overline{T}$ (see (\ref{eq:aveTE})
and (\ref{eq:aveTN})), which we aim to estimate, as a quantity alternative
to the (temporary) Hawking temperature $T_{\textrm{H}}$.

According to standard terminology, the notion \emph{radiometric} refers
to the quantities corresponding to radiant energy, whereas the notion
\emph{actinometri}c (\emph{photometric }or\emph{ photonic}) refers
to the quantities corresponding to the number of radiant particles
(photons) \cite{stewart2012blackbody}. In the case of spectral (i.e.,
$\omega$-dependent) quantities, the both possibilities are multiplicatively
related with the Planck multiplier $\hbar\omega$. The simplest relation
of this type is

\begin{equation}
e_{\omega}=\hbar\omega\mathcal{N}_{\omega},\label{eq:simplest}
\end{equation}
where $e_{\omega}$ is the \emph{spectral radiant exitance}, in the
case of the blackbody, given by the well-known \emph{Planck law} \cite{greiner1995thermodynamics}
(cf.\ \cite{Landau1969,Blundell2009})
\begin{equation}
e_{\omega}=\frac{\hbar\omega^{3}}{4\pi^{2}c^{2}}\cdot\frac{1}{e^{\nicefrac{\hbar\omega}{k_{\textrm{B}}T}}-1}\equiv\frac{\hbar\omega^{3}}{4\pi^{2}c^{2}}\textrm{Li}_{0}\left(e^{-\frac{\hbar\omega}{k_{\textrm{B}}T}}\right),\label{eq:eomega}
\end{equation}
and $\mathcal{N}_{\omega}$ is the corresponding \emph{spectral particle
}(\emph{photon})\emph{ exitance} assuming the standard form 
\begin{equation}
\mathcal{N}_{\omega}=\frac{\omega^{2}}{4\pi^{2}c^{2}}\cdot\frac{1}{e^{\nicefrac{\hbar\omega}{k_{\textrm{B}}T}}-1}\equiv\frac{\omega^{2}}{4\pi^{2}c^{2}}\textrm{Li}_{0}\left(e^{-\frac{\hbar\omega}{k_{\textrm{B}}T}}\right).\label{eq:calNomega}
\end{equation}
Here, $\hbar$ is the reduced Planck constant ($h/2\pi$), $c$ \textemdash{}
the speed of light, $k_{\textrm{B}}$ \textemdash{} the Boltzmann
constant, $\omega$ \textemdash{} the angular frequency, $T$ \textemdash{}
the blackbody temperature, and $\textrm{Li}_{0}$ is the polylogarithm
function $\textrm{Li}$ of the $0$th order (see Appendix). The notation
assumed in the paper is not quite standard \textemdash{} see Chapts.\ 34
and 36 in \cite{9780071629270} for standard notation.

Except where necessary, for clarity, we confine ourselves only to
radiometric quantities (actinometric quantities can be easily reproduced
according to (\ref{eq:rel_E_N_omega})). Nevertheless, we aim to treat
the both types of quantities equally and complementing each other.

The \emph{total radiant exitance}, denoted in our paper by $e$, is
given (for the blackbody) by the famous \emph{Stefan\textendash Boltzmann
law}
\begin{equation}
e=\int_{0}^{\infty}e_{\omega}d\omega=\frac{\pi^{2}}{60}\cdot\frac{k_{\textrm{B}}^{4}}{c^{2}\hbar^{3}}T^{4}\equiv\sigma T^{4},\label{eq:SBl}
\end{equation}
where $\sigma$ is the Stefan\textendash Boltzmann constant, whereas
the corresponding \emph{total particle (photon) exitance},
\begin{equation}
\mathcal{N}=\intop_{0}^{\infty}\mathcal{N}_{\omega}d\omega=\frac{\zeta\left(3\right)}{2\pi^{2}}\cdot\frac{k_{\textrm{B}}^{3}}{c^{2}\hbar^{3}}T^{3},\label{eq:tpe}
\end{equation}
with $\zeta\left(3\right)\approx1.20206$. For processes being considered
in a finite time interval, we can introduce the notion of the \emph{total
energy,} expressed by integration within this time interval,
\begin{equation}
E=\int Ae\thinspace dt,\label{eq:def_tot_energy}
\end{equation}
and the notion of the \emph{total number of particles},
\begin{equation}
N=\int A\thinspace\mathcal{N}\thinspace dt,\label{eq:def_tot_particles}
\end{equation}
where $A$ denotes the area of the surface of the source. Analogously,
the \emph{total spectral distribution of energy,} we are most interested
in, is defined by
\begin{equation}
E_{\omega}=\int Ae_{\omega}dt,\label{eq:def_N_omega}
\end{equation}
whereas the \emph{total spectral distribution of the number of particles}
defined by
\begin{equation}
N_{\omega}=\int A\mathcal{N}_{\omega}dt\label{eq:def_E_omega}
\end{equation}
is directly related to (\ref{eq:def_N_omega}) by the relation (cf.~(\ref{eq:simplest}))
\begin{equation}
E_{\omega}=\hbar\omega N_{\omega}.\label{eq:rel_E_N_omega}
\end{equation}

\section{Total number of particles\label{sec:Total number of particles}}

For the Schwarzschild BH, the spectral radiant exitance $e_{\omega}$,
and the spectral particle (photon) exitance $\mathcal{N}_{\omega}$
is given by the standard thermodynamic formulas (\ref{eq:eomega}),
and (\ref{eq:calNomega}), respectively, with $T$ equal the \emph{Hawking
BH temperature} $T_{\textrm{H}}$, i.e.,
\begin{equation}
T=T_{\textrm{H}}\equiv\frac{\hbar c^{3}}{8\pi k_{\textrm{B}}GM},\label{eq:TeqTH}
\end{equation}
where $G$ is the Newton gravitational constant.

Analogously, the total radiant exitance $e$, and the total particle
(photon) exitance $\mathcal{N}$ is given by another couple of standard
thermodynamic formulas, (\ref{eq:SBl}) and (\ref{eq:tpe}), respectively,
with $T=T_{\textrm{H}}$.

The total radiant energy (see (\ref{eq:def_tot_energy})) is directly
given by the Einstein mass\textendash energy equivalence formula 
\begin{equation}
E=Mc^{2},\label{eq:einstein}
\end{equation}
whereas the total number $N$ of particles (photons) emitted by the
BH has been determined quite recently in \cite{alonso2015entropy,muck2016hawking}.

Since, in the next section, for technical reasons, we will need in
our calculations the formula expressing the \emph{mass flow rate}
$dM/dt$ in terms of the mass $M$, involved in the derivation of
the evaporation time $t_{\textrm{e}}$, we rederive it in the present
section. As a byproduct of our derivation, we compute the total number
$N$ of particles (photons) emitted during the whole BH evaporation
process. By virtue of the equivalence formula (\ref{eq:einstein}),
and of the Stefan\textendash Boltzmann law (\ref{eq:SBl}), the mass
flow rate
\begin{equation}
\frac{dM}{dt}=-\frac{1}{c^{2}}\frac{dE}{dt}=-\frac{1}{c^{2}}Ae=-\frac{1}{c^{2}}A\sigma T^{4},\label{eq:dmpodt1}
\end{equation}
where $dE/dt$ can be interpreted as the \emph{velocity of evaporation}.
Inserting to (\ref{eq:dmpodt1}) the standard BH \emph{horizon surface
area formula}
\begin{equation}
A=\frac{16\pi G^{2}}{c^{2}}M^{2},\label{eq:defa}
\end{equation}
and the Hawking BH temperature formula (\ref{eq:TeqTH}), we obtain
the differential equation expressing the mass flow rate in terms of
$M$ 
\begin{equation}
\frac{dM}{dt}=-\frac{\hbar c^{4}}{15\cdot2^{10}\pi G^{2}}M^{-2}.\label{eq:dmpodt2}
\end{equation}
This equation yields the evaporation time
\begin{equation}
t_{\textrm{e}}=\frac{5\cdot2^{10}\pi G^{2}}{\hbar c^{4}}M_{0}^{3},\label{eq:tev}
\end{equation}
where $M_{0}$ denotes the initial mass of the BH.

As a byproduct of our derivation (\ref{eq:dmpodt2}), we can now calculate
the total number $N$ of particles (photons) emitted during the whole
BH evaporation process. To this end, we will apply a technical trick
which consists in using the chain rule to the \emph{number flow rate}
(given by differential form of (\ref{eq:def_tot_particles})), and
next performing some further formal manipulations (insertions). Thus,
first, we get
\begin{equation}
\frac{dN}{dt}=\frac{dN}{dM}\frac{dM}{dt}=A\mathcal{N}.\label{eq:dnpodt}
\end{equation}
Inserting the mass flow rate (\ref{eq:dmpodt2}) into the central
part of (\ref{eq:dnpodt}), and next the area formula (\ref{eq:defa}),
the total particle exitance (\ref{eq:tpe}), the Hawking temperature
(\ref{eq:TeqTH}) into the RHS of (\ref{eq:dnpodt}), respectively,
after simple rearrangements, we obtain, in agreement with earlier
derivations given in \cite{alonso2015entropy,muck2016hawking}, the
differential equation
\begin{equation}
\frac{dN}{dM}=-\frac{240\thinspace\zeta\left(3\right)G}{\pi^{3}\hbar c}M,\label{eq:dnpodm}
\end{equation}
where the Riemann zeta function $\zeta\left(3\right)\approx1.202$.
The solution of (\ref{eq:dnpodm}) yields the total number of particles
emitted
\begin{equation}
N=\frac{120\thinspace\zeta\left(3\right)G}{\pi^{3}\hbar c}M_{0}^{2}.\label{eq:obln}
\end{equation}
Thus, the total radiant energy $E$, and the total number $N$ of
particles (photons) emitted, both related to complete evaporation,
are determined.

\section{Total spectral distributions\label{sec:Total-spectral-distributions}}

The total spectral distribution $E_{\omega}$ of the total energy
$E$ emitted during the whole BH evaporation process is determined
by the \emph{spectral power} (given by differential form of (\ref{eq:def_N_omega}))
\begin{equation}
\frac{dE_{\omega}}{dt}=Ae_{\omega},\label{eq:Eomega1}
\end{equation}
or, equivalently, using the chain rule as a technical trick,
\begin{equation}
\frac{dE_{\omega}}{dM}=\left(\frac{dM}{dt}\right)^{-1}Ae_{\omega}.\label{eq:Eomega2}
\end{equation}
Making use of the mass flow rate formula (\ref{eq:dmpodt2}), the
area formula (\ref{eq:defa}) and the Planck law (\ref{eq:eomega}),
we obtain from (\ref{eq:Eomega2}) the \emph{(differential) element
of the total spectral distribution of energy}
\begin{equation}
dE_{\omega}=-\frac{15\cdot2^{12}G^{4}}{c^{10}}\omega^{3}M^{4}\textrm{Li}_{0}\left(e^{-\frac{8\pi G}{c^{3}}\omega M}\right)dM,\label{eq:dEomega1}
\end{equation}
or in terms of an (auxiliary) dimensionless BH mass
\begin{equation}
x\equiv\frac{8\pi G}{c^{3}}\omega M,\label{eq:defx}
\end{equation}
\begin{equation}
dE_{\omega}=-\frac{15c^{5}}{8\pi^{5}G}\omega^{-2}x^{4}\textrm{Li}_{0}\left(e^{-x}\right)dx.\label{eq:dEomega2}
\end{equation}
Now, we should integrate out the RHS of (\ref{eq:dEomega2}) with
respect to the mass $x$ over the finite mass interval $\left[x_{0},0\right]$,
corresponding to the total (initial) mass $M_{0}$ of the BH, where
\begin{equation}
x_{0}\equiv\frac{8\pi G}{c^{3}}\omega M_{0}\label{eq:dless_freq}
\end{equation}
is a dimensionless frequency. Utilizing the integral identity (\ref{eq:intxnLi0}),
we obtain our final formula for the total spectral distribution of
total energy
\begin{equation}
E_{\omega}=\frac{45c^{5}}{\pi^{5}G}\omega^{-2}\left[\zeta\left(5\right)-\sum_{k=0}^{4}\frac{x_{0}^{4-k}}{\left(4-k\right)!}\textrm{Li}_{k+1}\left(e^{-x_{0}}\right)\right].\label{eq:Eomega}
\end{equation}
Here, the Riemann zeta function $\zeta\left(5\right)$ ($\approx1.03693$)
is the only contribution from $0$ (the upper limit of the integral,
corresponding to the final mass $M_{\textrm{f}}=0$). More precisely,
for $k=4$ in the sum in the identity (\ref{eq:intxnLi0}) (see the
Riemann zeta function identity (\ref{eq:Lis1})),
\begin{equation}
\lim_{x\rightarrow0}\textrm{Li}_{4+1}\left(e^{-x}\right)=\textrm{Li}_{5}\left(1\right)=\zeta\left(5\right),\label{eq:limLi51}
\end{equation}
and there are no other contributions to (\ref{eq:intxnLi0}) from
0. In fact, for $k=1,2,3,$ we have (see (\ref{eq:Lis1}))
\begin{equation}
\lim_{x\rightarrow0}x^{4-k}\textrm{Li}_{k+1}\left(e^{-x}\right)=\left.x^{4-k}\zeta\left(k+1\right)\right|_{x=0}=0,\label{eq:limLikplus11}
\end{equation}
whereas for $k=0$ (see (\ref{eq:Li1}))
\begin{equation}
\lim_{x\rightarrow0}x^{4}\textrm{Li}_{1}\left(e^{-x}\right)=-\lim_{x\rightarrow0}x^{4}\ln\left(1-e^{-x}\right)=0.\label{eq:limLi1}
\end{equation}

\begin{figure}
\begin{centering}
\includegraphics[scale=0.1]{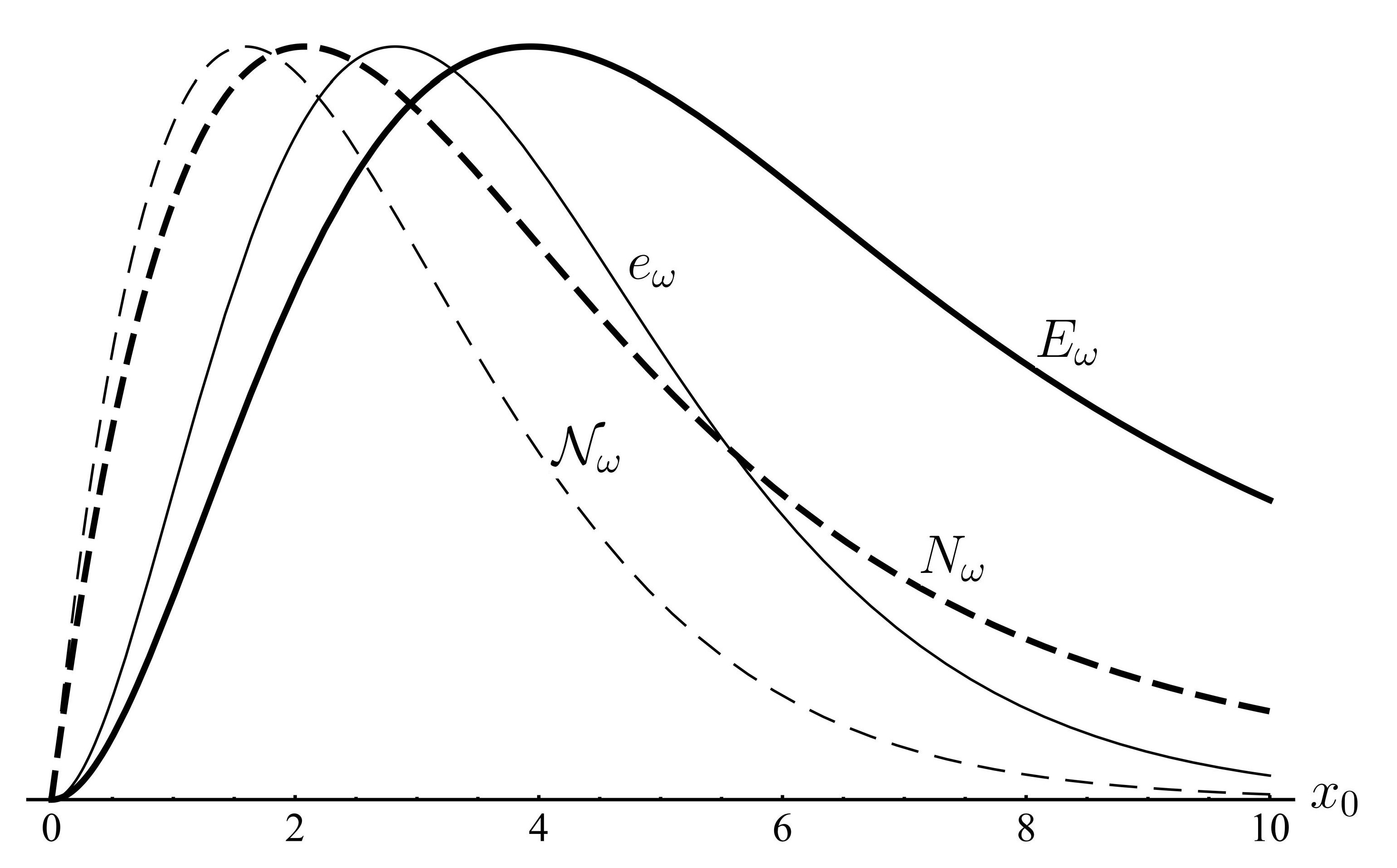}\medskip{}
\par\end{centering}
\begin{centering}
{\small{}}%
\begin{tabular}{rcccc}
\toprule 
{\small{}$F_{\omega}=$} & {\small{}$e_{\omega}$} & {\small{}$\mathcal{N}_{\omega}$} & {\small{}$E_{\omega}$} & \multirow{1}{*}{{\small{}$N_{\omega}$}}\tabularnewline
\midrule
{\small{}$x_{\textrm{max}}\left(F_{\omega}\right)=$} & {\small{}$2.82144$} & {\small{}$1.59362$} & {\small{}$3.93334$} & {\small{}$2.07408$}\tabularnewline
\bottomrule
\end{tabular}
\par\end{centering}{\small \par}
\centering{}\caption{\label{fig:Energy-and-particle}All the four spectral functions discussed
in the paper. Since the functions are expressed in different physical
units, they have been all normalized to a common ``one'', by convention,
and the vertical axis has been removed. Numerical values, denoted
by $x_{\textrm{max}}\left(F_{\omega}\right)$, of the dimensionless
frequency $x_{0}$ defined in (\ref{eq:dless_freq}), corresponding
to maxima of the spectral functions are given in the table above.}
\end{figure}

\section{Discussion and conclusions\label{sec:Conclusions}}

It seems that the best way to discuss the shape of the total spectral
distributions $E_{\omega}$ and $N_{\omega}$ for the BH, given by
(\ref{eq:Eomega}) and (\ref{eq:rel_E_N_omega}), is by comparison
to analogous temporary (initial) spectral exitances for the corresponding
blackbody, i.e., to $e_{\omega}$ and $\mathcal{N}_{\omega}$, respectively.
Differences and similarities between respective spectral functions
can be directly observed in Fig.~\ref{fig:Energy-and-particle}.
Their maxima, $x_{\textrm{max}}\left(F_{\omega}\right)$, have been
numerically determined, and they are given in the table in Fig.~\ref{fig:Energy-and-particle}.
In particular, we can observe that the values of the arguments for
the maxima of the total distributions are (a bit) greater than the
values of the arguments for the maxima of their temporary (initial)
blackbody counterparts, i.e., $x_{\textrm{max}}\left(E_{\omega}\right)>x_{\textrm{max}}\left(e_{\omega}\right)$
and $x_{\textrm{max}}\left(N_{\omega}\right)>x_{\textrm{max}}\left(\mathcal{N}_{\omega}\right)$.
Since the shapes of the total spectral distributions in their central
parts (in particular, we can ignore IR and UV tails, which are a bit
unphysical, see \cite{visser2015thermality}) are similar to their
blackbody counterparts, making use of the above mentioned shifts of
the maxima, we can introduce the approximate notion of the \emph{(time)
average(d) temperature} $\overline{T}$. More precisely, actually
we deal with two a bit differing average(d) temperatures. One temperature,
$\overline{T}_{E}$, corresponds to energy, and the second one, $\overline{T}_{N}$,
corresponds to the number of particles, respectively, and they both
are multiplicatively related to $T_{\textrm{H}}$. Namely,
\begin{equation}
\overline{T}_{E}\equiv\dfrac{x_{\textrm{max}}\left(E_{\omega}\right)}{x_{\textrm{max}}\left(e_{\omega}\right)}T_{\textrm{H}}\approx1.39409\cdot T_{\textrm{H}},\label{eq:aveTE}
\end{equation}
and

\begin{equation}
\overline{T}_{N}\equiv\dfrac{x_{\textrm{max}}\left(N_{\omega}\right)}{x_{\textrm{max}}\left(\mathcal{N}_{\omega}\right)}T_{\textrm{H}}\approx1.30149\cdot T_{\textrm{H}},\label{eq:aveTN}
\end{equation}
respectively. Here, the Hawking temperature $T_{\textrm{H}}$ is the
temporary blackbody temperature of the initial BH. Thus, on average,
the BH temperature is some $1.3$\textendash $1.4$ times greater
than its initial Hawking temperature $T_{\textrm{H}}$.

Since the Hawking radiation of a BH is a blackbody radiation with
the temperature $T_{\textrm{H}}$ which changes in time, and the BH
evaporation time $t_{\textrm{e}}$ is finite (see (\ref{eq:tev})),
in this paper, we have focused on the total spectral distributions
corresponding to the complete evaporation process, rather than on
temporary (initial) blackbody quantities (spectral exitances). The
total spectral distribution of energy, $E_{\omega}$, and the total
spectral distribution of the number of particles, $N_{\omega}$, have
been explicitly calculated and compared to their initial blackbody
counterparts.
\begin{acknowledgements}
Supported by the University of \L \'{o}d\'{z} grant.
\end{acknowledgements}

\appendix

\section*{Appendix}

The polylogarithm function (polylogarithm, in short) has already proved
to be very useful in the context of problems of conventional blackbody
radiation. In particular, exact formulas for blackbody radiation within
a given finite spectral band elegantly express in terms of polylogarithms
\cite{stewart2012blackbody}.

For real or complex $s$ and $z$ the \emph{polylogarithm} $\textrm{Li}_{s}(z)$
is defined by \cite{olver2010nist}

\begin{equation}
\textrm{Li}_{s}\left(z\right)\equiv{\displaystyle \sum_{k=1}^{\infty}\frac{z^{k}}{k^{s}}.}\label{eq:defLi}
\end{equation}
For each fixed complex $s$ the series defines an analytic function
of $z$ for $\left|z\right|<1$. The series also converges when $\left|z\right|=1$,
provided that $\Re s>1$. For other values of $z$, $\textrm{Li}_{s}(z)$
is defined by analytic continuation.

In particular,
\begin{equation}
\textrm{Li}_{0}\left(z\right)=\frac{z}{1-z},\label{eq:Li0}
\end{equation}
and
\begin{equation}
\textrm{Li}_{1}\left(z\right)=-\ln\left(1-z\right).\label{eq:Li1}
\end{equation}
The special case $z=1$ is the \emph{Riemann zeta function}: 
\begin{equation}
\zeta\left(s\right)=\textrm{Li}_{s}\left(1\right).\label{eq:Lis1}
\end{equation}
Integrating by parts, we can verify the following very useful in our
calculations integral identity \cite{stewart2012blackbody}
\begin{equation}
\int x^{n}\textrm{Li}_{0}\left(e^{-x}\right)dx=\textrm{const}.-n!\sum_{k=0}^{n}\frac{x^{n-k}}{\left(n-k\right)!}\textrm{Li}_{k+1}\left(e^{-x}\right),\label{eq:intxnLi0}
\end{equation}
valid for any non-negative integer $n$.

\bibliographystyle{spphys}
\bibliography{Total_spectral_distributions_from_Hawking_radiation}

\end{document}